\documentclass[pra,letterpaper,twocolumn,aps,footinbib,superscriptaddress,showpacs,babel]{revtex4-2}
\usepackage{amsmath,mathtools,upgreek,amssymb,graphicx,paralist,dsfont,color,transparent, float}
\usepackage[colorlinks = true, linkcolor = blue, urlcolor  = blue, citecolor = blue, anchorcolor = blue]{hyperref}
\usepackage{comment}
\usepackage{color}

\newcommand{\ket}[1]{|#1\rangle}
\newcommand{\bra}[1]{\langle#1|}
\newcommand{\braket}[1]{\langle#1\rangle}
\newcommand{\tr}{\mathrm{Tr}}
\newcommand{\dd}{\mathrm{d}}
\newcommand{\kett}[1]{|\negmedspace #1 \rangle}
\newcommand{\braa}[1]{\langle #1 \negmedspace|}

\begin{document}

\title{GHZ protocols enhance frequency metrology despite spontaneous decay}

\author{T. Kielinski}
\affiliation{Institute for Theoretical Physics and Institute for Gravitational Physics (Albert-Einstein-Institute), Leibniz University Hannover, Appelstrasse 2, 30167 Hannover, Germany}

\author{P. O. Schmidt}
\affiliation{Institute of Quantum Optics, Leibniz University Hannover, Welfengarten 1, 30167 Hannover, Germany}
\affiliation{Physikalisch-Technische Bundesanstalt, Bundesallee 100, 38116 Braunschweig, Germany}

\author{K. Hammerer}
\email{klemens.hammerer@itp.uni-hannover.de}
\affiliation{Institute for Theoretical Physics and Institute for Gravitational Physics (Albert-Einstein-Institute), Leibniz University Hannover, Appelstrasse 2, 30167 Hannover, Germany}

\begin{abstract}
The use of correlated states and measurements promises improvements in the accuracy of frequency metrology and the stability of atomic clocks. However, developing strategies robust against dominant noise processes remains challenging. We address the issue of decoherence due to spontaneous decay and show that Greenberger-Horne-Zeilinger (GHZ) states, in conjunction with a correlated measurement and nonlinear estimation strategy, achieve gains of up to 2.25 dB, comparable to fundamental bounds for up to about 80 atoms in the presence of decoherence. This result is surprising since GHZ states do not provide any enhancement under dephasing due to white frequency noise compared to the standard quantum limit of uncorrelated states. The gain arises from a veto signal, which allows for the detection and mitigation of errors caused by spontaneous emission events. Through comprehensive Monte-Carlo simulations of atomic clocks, we demonstrate the robustness of the GHZ protocol.
\end{abstract}

\maketitle

Frequency metrology represents a cornerstone of modern precision measurements, playing a crucial role in driving advancements and fostering progress across numerous research areas in physics~\cite{Ludlow2015,Pezz2018,Colombo2022}. In particular, optical atomic clocks are the most precise measurement devices achieving stabilities on the scale of $10^{-18}$ and below~\cite{Guideline2019,Oelker2019,Schioppo2016,Nicholson2015,Bloom2014,PedrozoPeafiel2020,Aepelli2024,Keller2019,Hausser2024,Aharon2019,Sanner2019,Huang2017}. This unprecedented performance opens the path towards new applications in fundamental research and technology, such as the redefinition of the SI second~\cite{Ludlow2015}, relativistic geodesy~\cite{Lisdat2016,Grotti2018,Mehlstubler2018,Grotti2024}, tests of general relativity~\cite{Takamoto2020,Chou2010,Bothwell2022,Dimarcq2024,Dreissen2022,Sanner2019} and the search for physics beyond the standard model~\cite{Derevianko2014,Safronova2019,Roberts2020}. 
    
Current efforts to further improve the stability of optical clocks involve exploring the use of entanglement in atomic systems to reduce quantum projection noise and overcome the standard quantum limit (SQL) imposed by uncorrelated atoms~\cite{Ludlow2015,Pezz2018,Colombo2022}. This is also spurred by the development of highly-controlled platforms for optical clocks based on tweezer arrays~\cite{madjarov2019,norcia2019,young2020,shaw2024} and Coulomb crystals~\cite{steinel2023,pelzer2023,Hausser2024} which allow for engineered correlations among atoms. Recent progress demonstrating entanglement on optical clock transitions includes the generation of spin squeezing in trapped ions~\cite{franke2023} and in neutral atoms mediated via cavities~\cite{PedrozoPeafiel2020,robinson2024} or Rydberg interactions~\cite{Eckner2023}. Recently, also GHZ states and cascades thereof have been shown in optical clocks based on tweezer arrays~\cite{Finkelstein2024,Cao2024}.

Unfortunately, decoherence presents a substantial obstacle in frequency metrology, impairing the precision of measurements by compromising the coherence of quantum systems essential for achieving entanglement-based enhancement~\cite{Shaji2007,UlamOrgikh2001,Demkowicz_Dobrza_ski_2012,Escher2011,Huelga1997,Sekatski2017}. Among technical noise sources, residual fluctuations of the laser and magnetic field noise impose major limitations~\cite{Ludlow2015,Pezz2018,Colombo2022, Leroux2017,Guideline2019,Monz2011}. In particular, it has been demonstrated that GHZ protocols, which are optimal in the absence of decoherence, are highly sensitive to dephasing due to frequency fluctuations and ultimately show no improvement over SQL~\cite{Huelga1997}. To address this susceptibility of entanglement-enhanced protocols, frequency fluctuations were taken into account to determine optimal interrogation sequences~\cite{Kaubruegger2021,Thurtell2022,Schulte2020} as well as multi-ensemble schemes to effectively extend the coherence time were proposed recently~\cite{Hume2016,Rosenband2013,Cao2024,Nardelli21,Borregaard2013,Drscher2020,Li2022,Zheng2024}.

\begin{figure}[tbp]
        \centering
            \includegraphics[width=\linewidth]{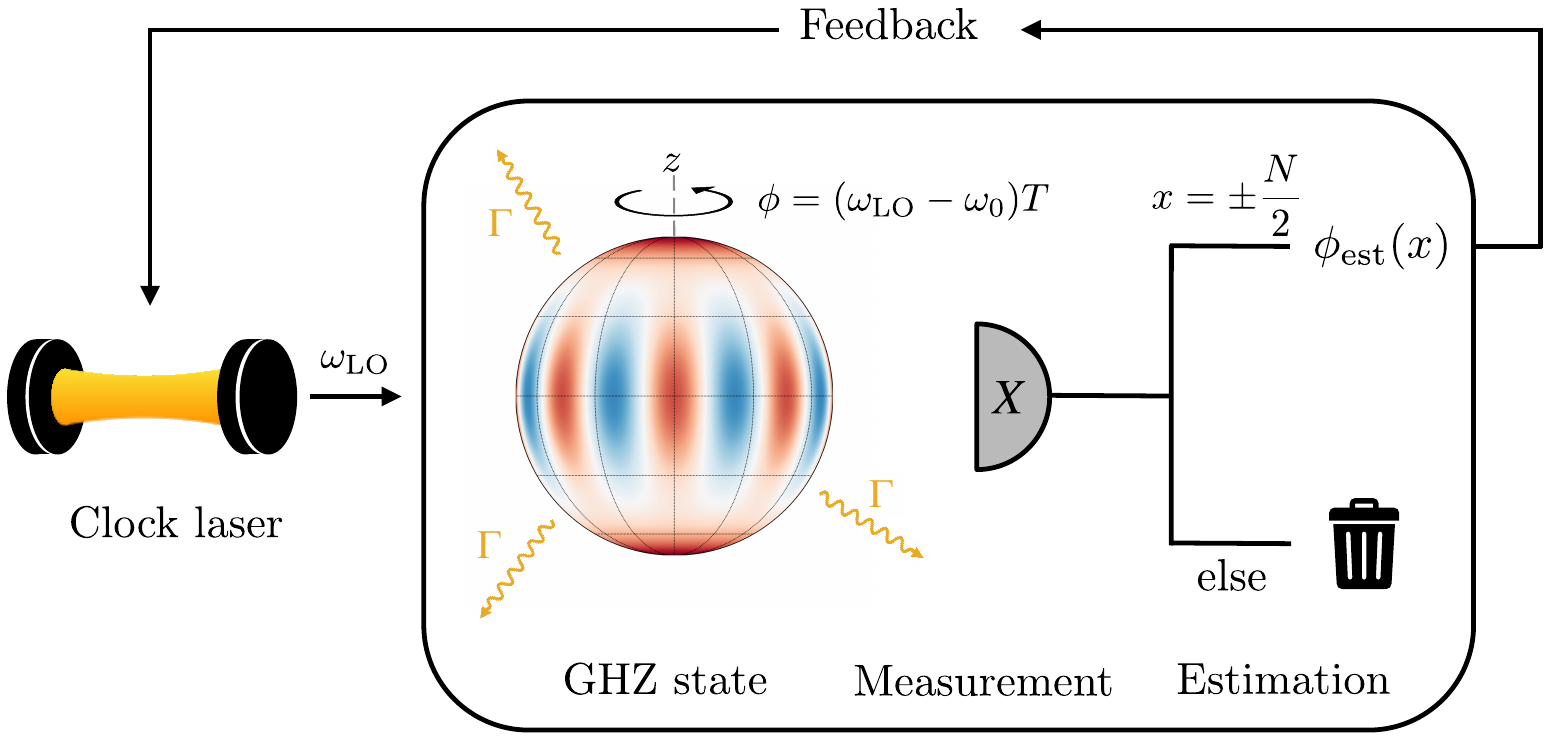}
     
        \caption{\textbf{Optimal GHZ protocol embedded in the full feedback loop of an atomic clock.} A local oscillator (LO) with fluctuating frequency $\omega_\mathrm{LO}$ is stabilized to an atomic transition $\omega_0$. During the free evolution time $T$, the probe state accumulates a phase $\phi$ and is subject to spontaneous emission with rate $\Gamma$. Based on the measurement outcome $x$ of the observable $X$, an estimation $\phi_\mathrm{est}$ of the phase is conducted to correct the LO accordingly.}
        \label{fig:Scheme}
    \end{figure}

In contrast to the extensive treatment of frequency fluctuations, the effects of spontaneous emission have received comparatively little attention. However, the finite lifetime of the qubits in the excited state represents a fundamental limit rather than an external noise source. With state-of-the-art clock lasers reaching coherence times of several seconds~\cite{Matei2017}, the excited-state lifetime of various clock candidates becomes the more stringent limitation, for instance Ca$^+$-ions ($1.1\,\mathrm{s}$), Sr$^+$-ions ($0.4\,\mathrm{s}$), In$^+$-ions ($0.2\,\mathrm{s}$) and Hg-atoms ($1.6\,\mathrm{s}$). With further technological improvements in the short-term stabilization of lasers, even clock species as Al$^+$-ions ($20.7\,\mathrm{s}$) and Yb-atoms ($15.9\,\mathrm{s}$) will enter this regime~\cite{Guideline2019}. This applies especially when using strongly entangled states which are generally more susceptible to decoherence e.g. due to spontaneous emission. Consequently, it is highly relevant to investigate the impact of spontaneous decay for the development of future clocks and identify optimal interrogation schemes for particular setups.

Here, we will present a protocol with quantum operations of low complexity and a highly nonlinear estimator that saturates the quantum Cramér-Rao bound (QCRB) of the GHZ state (cf.~\autoref{fig:Scheme}). Surprisingly, and in contrast to dephasing, we find that GHZ states provide a substantial enhancement compared to the SQL in the presence of spontaneous decay. Moreover, we will compare the sensitivity of this protocol to the ultimate lower limit~\cite{Macieszczak2013} and to squeezed spin states (SSS) which are optimal in the asymptotic limit for large particle number. In addition, we will present a variation of this protocol with a GHZ-like initial state reaching the level of the ultimate lower limit for up to about 80 atoms. Finally, we will show the robustness of the measurement and estimation scheme by performing Monte-Carlo simulations of the full feedback loop in an atomic clock.  Note that although we focus on the specific application to atomic clocks, the underlying concept extends to Ramsey interferometry in general, as utilized in various precision measurements, such as atom interferometry and magnetometry.

To implement the low complexity protocol, we employ single one-axis-twisting (OAT)~\cite{Kitagawa1993} operations for both state preparation and as an effective measurement, since they give rise to a variety of entangled states, ranging from squeezed spin states (SSS) to the GHZ state, as well as variational classes of generalized Ramsey spectroscopy saturating the ultimate limit in sensitivity~\cite{SchulteEcho2020,Scharnagl2023,Kaubruegger2021,Thurtell2022,Marciniak2022,Li2023}. Furthermore, OAT interactions are accessible in several setups as in ion traps via Mølmer-Sørensen gates~\cite{Benhelm2008,Blatt2008,Leibfried2004}, in tweezer arrays via Rydberg interactions~\cite{Eckner2023,Hines2023} or Bose-Einstein condensates via elastic collisions~\cite{Srensen2001,Estve2008,Gross2010,Riedel2010}.

\paragraph*{GHZ protocols in frequency metrology ---} We start by summarizing the main idea and problems associated with improving frequency metrology using GHZ entanglement. In an atomic clock, the frequency deviation $\omega = \omega_{\mathrm{LO}} - \omega_0$ of the local oscillator $\omega_{\mathrm{LO}}$ from an atomic reference transition $\omega_0$ is estimated by repeated Ramsey interrogations measuring a phase $\phi = \omega T$, where $T$ is the interrogation time. Averaging over $n$ independent and identical interrogations achieves an uncertainty in frequency estimation given by 
\begin{align}
    \Delta\omega(T) = \frac{\Delta\phi(T)}{\sqrt{T\tau}}\label{eq:freqUncertainty}
\end{align}
where we fix the total averaging time $\tau = nT$ and denote $\Delta\phi(T)$ as the phase estimation uncertainty of a single Ramsey interrogation (cf. Methods). Disregarding decoherence, $\Delta\phi$ is independent of time, and the frequency estimation uncertainty appears to monotonically decrease for longer interrogation times $T$. If $N$ atoms are prepared in uncorrelated states, the phase uncertainty is limited by the standard quantum limit (SQL) $\Delta\phi^2 \geq \Delta\phi_\mathrm{SQL}^2=1/N$~\cite{Pezz2018}, such that $\Delta\omega(T)^2 \geq (\Delta\omega_\mathrm{SQL}(T))^2=1/\tau T N$. This bound can be saturated with atoms prepared in a coherent spin state (CSS) and measurement of the spin projection along a suitable direction. A dramatic improvement can, in principle, be gained by preparing atoms in a GHZ state $\ket{\mathrm{GHZ}} = \left(\kett{\uparrow}^{\otimes N} + \kett{\downarrow}^{\otimes N}\right)/\sqrt{2}$ and measuring the parity $\Pi = (-1)^N\sigma_x^{\otimes N}$ after phase imprint during a Ramsey interrogation, as initially proposed by Wineland et al.~\cite{Bollinger1996}. This strategy, to which we refer in the following as `parity-GHZ' protocol, surpasses the SQL and saturates the Heisenberg limit (HL) $(\Delta\phi_{\mathrm{HL}})^2 = 1/N^2$~\cite{Pezz2018}. The improvement is due to the signal in Ramsey interference oscillating $N$ times faster, thus enhancing the sensitivity by exactly this factor.

Unfortunately, this figure is severely compromised by decoherence~\cite{Shaji2007,Huelga1997,UlamOrgikh2001,Demkowicz_Dobrza_ski_2012,Escher2011,Sekatski2017}. In particular, a loss of coherence means that $\Delta\phi(T)$ increases with $T$, so that a compromise must be found for the optimal interrogation time $T_\mathrm{min}$ with regard to the frequency uncertainty in Eq.~\eqref{eq:freqUncertainty}. Considering individual dephasing and spontaneous decay of atoms with rates $\gamma$ and $\Gamma$ respectively, one finds the SQL for uncorrelated atoms (cf. Methods)
\begin{align}
    (\Delta\omega_\mathrm{SQL})^2 = \frac{\mathrm{e}\,(\Gamma + \gamma)}{N\tau}\label{eq:sql}
\end{align}
which is achieved at an optimal interrogation time $T^\mathrm{min}_\mathrm{SQL} = 1/(\Gamma+\gamma)$. This is consistent with the bound derived for vanishing spontaneous emission rate $\Gamma= 0$ by Huelga et al.~\cite{Huelga1997}, where it is also demonstrated that parity-GHZ protocols completely lose their gain due to dephasing, achieving at best the frequency uncertainty given by Eq.~\eqref{eq:sql}. Extending the analysis to additionally include spontaneous decay yields analogous observations (cf. Methods). This outcome results from GHZ states decohering $N$ times faster than uncorrelated states, leading to a proportionally shorter optimal interrogation time, exactly compensating for the gain in phase uncertainty. As a consequence, this statement has often been generalized without further investigation in the sense that GHZ states generally do not lead to any improvement in the presence of decoherence. 
    
\begin{figure}[tbp]
        \centering
            \includegraphics[width=\linewidth]{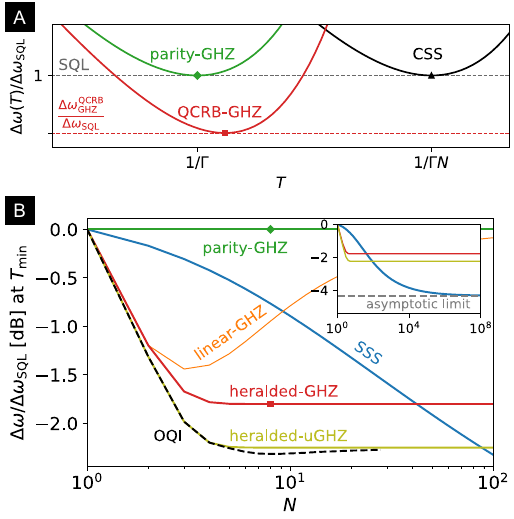}
     
        \caption{\textbf{Enhancement of the optimal GHZ protocol compared to the SQL.} (\textbf{A}) Generic frequency estimation uncertainty $\Delta\omega(T)$ in the presence of spontaneous decay for $N=8$ scaled to the SQL~\eqref{eq:sql}, $\gamma=0$. Symbols indicate the optimal interrogation time $T_\mathrm{min}$. (\textbf{B}) Relative frequency estimation uncertainty at the optimal interrogation time $T_\mathrm{min}$ for the indicated protocol types (see main text). The ratio $\Delta\omega/\Delta\omega_{\mathrm{SQL}}$ is independent of the decay rate $\Gamma$ due to the comparison with the SQL (triangle in (a)). Square and diamond symbols correspond to the minima in (a). The inset illustrates the asymptotic scaling for large ensembles $N$, using the same axes and protocols.}
        \label{fig:Theory}
    \end{figure}

\paragraph*{Bounds for uncertainty with GHZ states---} For a given probe state, the quantum Cram\'{e}r-Rao bound (QCRB) represents a lower limit in phase estimation, i.e. depicts the minimal phase uncertainty with respect to all conceivable measurement and estimation schemes \cite{Pezz2018}. Indeed, performing a parity measurement on the GHZ state in the presence of spontaneous decay is not optimal, i.e. does not saturate the QCRB. We show in the Supplementary Materials that the QCRB for the phase uncertainty is given by 
    \begin{align}
        (\Delta\phi_\mathrm{QCRB}^\mathrm{GHZ}(T))^2  = \frac{ \mathrm{e}^{(\Gamma + \gamma )NT}}{2N^2} \left[1 + e^{-\Gamma N T}+\left(1-e^{-\Gamma T}\right)^N\right].\label{eq:GHZQCRB}
    \end{align}
The corresponding bound for the optimal frequency uncertainty follows from minimizing with respect to the interrogation time, $(\Delta\omega_\mathrm{QCRB}^\mathrm{GHZ})^2=\min_T (\Delta\phi_\mathrm{QCRB}^\mathrm{GHZ}(T))^2/T\tau$. For vanishing spontaneous emission, $\Gamma=0$, this reproduces the result of Huelga et al.~\cite{Huelga1997} that the QCRB equals the SQL. However, for $\Gamma> 0$ one finds $(\Delta\omega_\mathrm{QCRB}^\mathrm{GHZ})^2<(\Delta\omega_\mathrm{SQL})^2$ which is illustrated in \autoref{fig:Theory}A. Somewhat surprisingly, GHZ states do admit gains beyond the SQL when the relevant decoherence process is spontaneous emission rather than dephasing noise. Since these gains are not realized by parity measurements, the question arises as to which alternative measurements do saturate the QCRB of the GHZ state \eqref{eq:GHZQCRB}.

We show in the Supplementary Materials that this can be achieved as follows (cf.~\autoref{fig:Scheme}): At the end of a Ramsey interrogation, atoms are subject to a unitary operation
\begin{align}
    \mathcal{U}_\mathrm{GHZ} = \begin{cases}
        \mathcal{T}_x(\pi) & \text{if }N\text{ is even}\\
        \mathcal{R}_x\left(\frac{\pi}{2}\right)\mathcal{T}_x(\pi) & \text{if }N\text{ is odd}
    \end{cases}\label{eq:GHZGHZ}
\end{align}
where $\mathcal{T}_x(\mu) = \exp\left(-i\frac{\mu}{2}S_x^2\right)$ denotes an OAT operation with twisting strength $\mu=\pi$ along $x$, whereas $\mathcal{R}_x(\theta)$ represents a rotation around the $x$-axis by an angle $\theta=\frac{\pi}{2}$. We note that the unitary $\mathcal{U}_\mathrm{GHZ}$ also corresponds to the operation that may be used to generate the GHZ state from the ground state $\kett{\!\downarrow}^{\otimes N}$ initially. Subsequently, atoms are measured projectively along $z$, effectively measuring the operator $X = \mathcal{U}_\mathrm{GHZ} \,S_z\,\mathcal{U}_\mathrm{GHZ}^\dagger$ with outcomes $x\in\{-\frac{N}{2},\ldots,\frac{N}{2}\}$. This procedure corresponds to the one explored in the experiment reported by Leibfried et al. \cite{Leibfried2004} and essentially implements an exact Loschmidt echo \cite{LoschmidtEcho}, since the state preparation with $\mathcal{U}_\mathrm{GHZ}$ is complemented by the corresponding adjoint transformation $\mathcal{U}_\mathrm{GHZ}^\dagger$ for the measurement. Any such measurement has to be accompanied by a suitable rule for estimating the phase $\phi$ for a given outcome $x$. We can show that the nonlinear estimator
\begin{align}
    \phi_\mathrm{est}(x,T) =\begin{cases}
          \pm \mathrm{e}^{\frac{(\Gamma + \gamma)}{2}NT} &\text{if }x =\pm\frac{N}{2}\\
        0&\text{else}
    \end{cases}\label{eq:flag}
\end{align}
has to be applied in order to saturate the QCRB. $\phi_\mathrm{est}$ estimates the phase for the maximal outcomes $x=\pm \frac{N}{2}$ according to the standard linear estimator in local phase estimation (cf. Methods), namely linearly scaled with the inverse signal slope, while all other measurement outcomes are simply neglected. In the context of atomic clocks, an estimated phase $\phi_\mathrm{est} = 0$ corresponds to inferring no frequency deviation, and correspondingly, no error signal is generated.

The fact that this measurement and estimation strategy performs well under spontaneous emission can be understood from the two following reasons: Firstly, measurement outcomes $x \neq \pm\frac{N}{2}$ can occur  only if at least one spontaneous emission event happened, such that selecting the $x = \pm\frac{N}{2}$ events, as is done by the estimator~\eqref{eq:flag}, in essence filters out all such cases. Secondly, and quite remarkably, for the particular measurement $X$ introduced above, the conditional probabilities $P(x|\phi,T)$ turn out to be independent of $\phi$ for all $x \neq \pm\frac{N}{2}$ (cf. Methods). Thus, these cases don't deliver any information on $\phi$. Together, these two features allow to implement an error detection and mitigation scheme~\cite{Cai2023} tailored to frequency metrology. On the one hand, outcomes $x \neq \pm\frac{N}{2}$ can be considered as heralded errors signifying an unsuccessful, decohered Ramsey interrogation. Conversely, outcomes $x = \pm\frac{N}{2}$ signal a no-jump dynamics (cf. Supplementary Materials) delivering maximal phase information. On the other hand, exclusively selecting these events during (classical) post-processing results in enhanced sensitivity. We therefore refer to this scheme as `heralded-GHZ' protocol in the following. 

Two immediate concerns arise regarding this protocol. Firstly, how well does it compare to other potential strategies where, beyond measurements and estimators, the initial state is also optimally chosen and may differ from a GHZ state? Secondly, one may question the effectiveness of a strategy that ignores all but two measurement outcomes in each interrogation cycle, particularly in the context of atomic clocks, where the ultimate challenge is to stabilize the constantly drifting phase of a local oscillator. We address both of these issues in the remainder of this article.

\paragraph*{Comparison to optimal quantum interferometer---} Optimization over all entangled initial states, measurements, and estimators determines the performance of an optimal quantum interferometer (OQI). In the absence of decoherence, the OQI achieves the Heisenberg limit of sensitivity. Unfortunately, in the presence of decoherence, no general expressions for the OQI sensitivity for arbitrary ensemble sizes are available, but they rather require complex optimization procedures. An algorithm presented in Ref.~\cite{Macieszczak2013} iteratively optimizes the initial probe state and the observable, which can be done efficiently at least for up to $N=30$ atoms. However, numerical optimization becomes challenging with increasing ensemble size $N$ (cf. Supplementary Materials). In the asymptotic limit ($N \gg 1$), lower bounds can be derived for several noise processes as discussed in Refs.~\cite{Escher2011,Demkowicz_Dobrza_ski_2012,Huelga1997,Kurdziaek2023,DemkowiczDobrzaski2014,Knysh2014}. The ultimate lower bound in the presence of dephasing~\cite{Huelga1997,Escher2011,Demkowicz_Dobrza_ski_2012} and spontaneous decay~\cite{Knysh2014} is given by
    \begin{align}
        (\Delta\omega_\mathrm{min}^\mathrm{asym})^2 \geq \frac{\Gamma + \gamma}{N \tau} \label{eq:asymptotic}
    \end{align}
    which yields a maximal improvement of $1/\mathrm{e}$ over the SQL~\eqref{eq:sql}, corresponding to a gain of 4.3 dB, and is asymptotically saturated by SSS (cf. Supplementary Materials). As a consequence, we will focus on small ensembles where the OQI can be evaluated numerically and benchmark the heralded-GHZ protocol against SSS generated by OAT since they are optimal in the asymptotic limit. In the following, we will consider setups where spontaneous decay is the main limiting decoherence process, and neglect dephasing noise assuming $\gamma=0$. 
    
    In \autoref{fig:Theory}B, we present the sensitivity of frequency estimation for a given particle number $N$, optimized in each case with respect to the interrogation time $T$ and scaled to the SQL. The conclusions we draw from~\autoref{fig:Theory}B are: (i) The heralded-GHZ protocol shows a substantial enhancement over the SQL for all $N$. In particular, a constant gain of 1.8 dB with regard to the the SQL is achieved for $N\geq 4$ and thus, no loss in improvement is observed for large $N$. Especially considering the scaling of small and intermediate ensembles, the resulting increase in sensitivity is quite remarkable. (ii) The importance of the nonlinear estimator~\eqref{eq:flag} becomes evident since otherwise (denoted as `linear-GHZ' protocol), i.e. using the standard linear estimator,  the maximal gain is obtained for $N=3$, whereas for larger ensembles the sensitivity converges to the SQL (cf. Methods). (iii) Although the heralded-GHZ protocol provides a substantial enhancement compared to the SQL, it does not saturate the OQI, which likewise represents a constant improvement with respect to the SQL in the regime ($N\leq 30$) where the numerical evaluation is feasible (cf. Supplementary Materials). Furthermore, in this regime the heralded-GHZ protocol achieves sensitivities relatively close to the OQI, despite its low complexity. (iv) For larger ensembles ($N\geq 42$), SSS achieve a higher sensitivity than the heralded-GHZ protocol and ultimately approximate the lower bound~\eqref{eq:asymptotic} asymptotically. Consequently, the gain of the OQI over the SQL will increase likewise for larger ensembles. In summary, this implies that GHZ states can be used to overcome the SQL in the presence of spontaneous emission and in addition perform close to the OQI for small and moderate ensemble sizes.

    \paragraph*{Saturating the OQI---} The gap between the heralded-GHZ protocol and the OQI (cf.~\autoref{fig:Theory}B) raises the question of which protocol could be used to close it and what resources would be required to do so. The fact that the gap is independent of the number of particles suggests that this could be possible with a fixed protocol that varies little or not at all with $N$. Indeed, we have identified a particular interrogation scheme that reaches the level of the OQI for up to about 80 atoms. Surprisingly, no deep circuit depths are necessary, but a simple extension of the heralded-GHZ protocol involving one more twisting operation for state preparation is sufficient. The initial state
    \begin{align}
        \ket{\psi_\mathrm{in}} &= \mathcal{U}_\mathrm{GHZ} \,\mathcal{R}_z(\theta)  \,\mathcal{U}_\mathrm{GHZ}\kett{\downarrow}^{\otimes N}=\alpha_\theta\kett{\downarrow}^{\otimes N}+\beta_\theta\kett{\uparrow}^{\otimes N} \label{eq:unbalancedGHZstate}
    \end{align}
    is generated by an additional OAT interaction $\mathcal{U}_\mathrm{GHZ}$ and rotation $\mathcal{R}_z(\theta)$ with optimal rotation angle $\theta$ depending on the ensemble size $N$ and the dimensionless parameter $\Gamma T$ (cf. Supplementary Materials). Essentially, the additional transformation only modifies the coefficients $\alpha_\theta$ and $\beta_\theta$ generating an unbalanced GHZ-like state (referred to as uGHZ in the following). In particular, the optimal rotation angle gives higher weight to the excited state, compensating for decay during the free evolution time. Due to its GHZ-like nature, the same measurement \eqref{eq:GHZGHZ} and estimation scheme \eqref{eq:flag} are optimal as for the heralded-GHZ protocol and thus result in similar properties. We show in the Supplementary Materials that the corresponding phase estimation uncertainty reads
    \begin{align}
        (\Delta\phi(T))^2 = \frac{e^{(\Gamma + \gamma)N  T}}{4N^2 } \left[1 + \sqrt{e^{-N\Gamma T} + \left(1-e^{-\Gamma T}\right)^N}\right]^2.\label{eq:unbalancedGHZ}
    \end{align}
    \autoref{fig:Theory}B shows that this `heralded-uGHZ' protocol saturates the OQI for small ensemble sizes ($N\leq 4$) and achieves a constant gain of 2.25 dB compared to the SQL for $N\geq 6$, which is close to the OQI for intermediate $N$. Again, the asymptotically optimal SSS are advantageous for larger ensembles ($N>87$).

\paragraph*{Performance in atomic clock---} To investigate the robustness of the presented measurement and estimation schemes to spontaneous decay in a realistic scenario of frequency metrology, we perform numerical Monte Carlo simulations of the full feedback loop in an atomic clock and compare the results to the theoretical predictions. At the core, an atomic clock comprises a local oscillator (LO) with an inherently noisy frequency signal $\omega_\mathrm{LO}$ that varies over time. The frequency $\omega_\mathrm{LO}$ is stabilized to the atomic transition frequency $\omega_0$ by repeatedly interrogating the atomic ensemble (cf.~\autoref{fig:Scheme}). The control cycle is completed by the servo, which applies feedback based on the phase estimate $\phi_\mathrm{est}(x,T)$. The Monte Carlo simulation implements the basic principles of an atomic clock as described above and builds upon fundamental routines developed in Refs.~\cite{Leroux2017,Schulte2020,Kaubruegger2021}.

\begin{figure}[tbp]
        \centering
            \includegraphics[width=\linewidth]{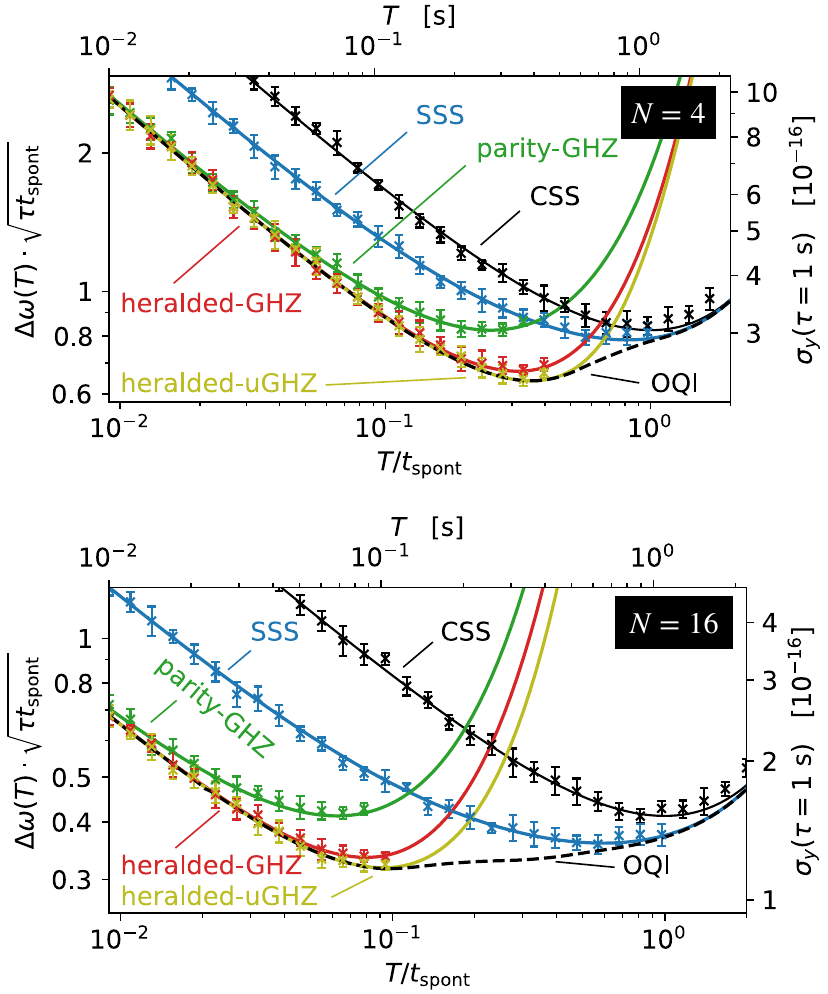}
     
        \caption{\textbf{Monte-Carlo simulations of the full feedback loop in an atomic clock.} Simulation results are displayed by symbols while theoretical predictions are shown as lines. Data points denote the average over $10$ independent clock runs. Error bars indicate the corresponding standard deviations. In each clock run, $10^7$ cycles were performed and the corresponding Allan deviation evaluated. The presented values are obtained by extrapolating the Allan deviation at $\tau=1~\mathrm{s}$ based on the long term stability~\eqref{eq:adev} at $\tau\gg 1$.}
            \label{fig:simulation}
    \end{figure}

    The long term stability of an atomic clock is quantified by the Allan deviation $\sigma_y(\tau)$~\cite{Allan1966} characterizing the fluctuations of fractional frequency deviations $y = \omega/\omega_0$ averaged over $\tau\gg T$. In a dead-time free scenario, the Allan deviation is well approximated by
    \begin{align}
        \sigma_y(\tau) = \frac{1}{\omega_0}\frac{\Delta\phi(T)}{\sqrt{T\tau}}=\frac{\Delta\omega(T)}{\omega_0}\label{eq:adev}
    \end{align}
    and thus corresponds to the frequency estimation uncertainty rescaled with $\omega_0$. Consequently, we present our results in two ways: Lower $x$-axis and left $y$-axis refer to general frequency estimation where the uncertainty is rescaled to be independent of the particular averaging time $\tau$ and lifetime $t_\mathrm{spont}=1/\Gamma$. Thus, it allows for an application to several experimental setups and atomic species as long as spontaneous decay remains the dominating decoherence effect. Upper $x$-axis and right $y$-axis illustrate results in an atomic clock for the particular example of Ca$^+$-ions with lifetime $t_\mathrm{spont} = \frac{1}{\Gamma} \simeq 1.1\,\mathrm{s}$~\cite{Gudjons1996} and transition frequency $\omega_0 = 2\pi\nu_0 \simeq  411.042\,\mathrm{THz}$~\cite{Huang2017}. Furthermore, we consider frequency fluctuations corresponding to a state of the art clock laser. In particular, we assume a laser limited by flicker noise with coherence time $Z\simeq 7.5~\mathrm{s} \gg t_\mathrm{spont}$~\cite{Matei2017}. Results of numerical simulations (symbols) in comparison to theoretical predictions (lines) of the investigated protocols are shown in~\autoref{fig:simulation} for $N=4$ (top) and $N=16$ (bottom) particles. As Monte Carlo simulations are stochastic processes, resulting stabilities fluctuate around the average value. Overall, numerical simulations of all discussed interrogation schemes show very good agreement with theoretical predictions. Therefore, all schemes, including the heralded-GHZ and heralded-uGHZ protocols in particular, are robust and thus suited for realistic scenarios as in the context of atomic clocks. At long interrogation times $T\gtrsim t_\mathrm{spont}$, fringe hops occur where the feedback loop passes to an adjacent fringe resulting in the clock running systematically wrong and consequently spoiling the clock stability. Moreover, the comparison of $N=4$ and $N=16$ already indicates the transition between the optimality of GHZ-like protocols and squeezing protocols.

\paragraph*{Conclusion \& Outlook---}
We have presented a protocol with low complexity that saturates the QCRB of GHZ states and thus, unexpectedly, results in a substantial enhancement of 1.8 dB compared to the SQL in the presence of spontaneous decay. This is achieved by a measurement and estimation scheme that allows to identify and exclude spontaneous emission events in the Ramsey sequence and thus implements an error detection and mitigation scheme to improve frequency metrology. Additionally, we have identified a GHZ-like protocol saturating the OQI for small ensembles and performs closely to it for intermediate ensemble sizes. The 2.25 dB gain compared to the SQL results from an unequal superposition of the two maximal Dicke states, which counteracts spontaneous decay during the free evolution time and can be generated by a minor modification of the GHZ state. Furthermore, the robustness of these protocols was shown through comprehensive Monte-Carlo simulations of atomic clocks, thereby paving the way for near-term implementations into experimental setups. GHZ(-like) states are an attractive experimental choice, since in a variation of the protocol involving the entanglement of Zeeman states of equal magnitude but opposite sign, they can be made first order magnetic field insensitive~\cite{Roos2006}, thus eliminate  dephasing, the major source of decoherence~\cite{Monz2011}. Furthermore, the shorter optimum probe time compared to CSS reduces loss of contrast and time dilation shifts from motional heating of the ion crystal, thus improving the signal-to-noise ratio and the accuracy of such a clock. An open question remains to investigate the transition from GHZ-like protocols for small ensembles to SSS in the asymptotic limit. In addition to frequency metrology, the findings of this work can be applied to various precision measurements where phase estimation is limited by spontaneous decay.

\section*{Acknowledgments}
We acknowledge funding by the Deutsche Forschungsgemeinschaft (DFG, German Research Foundation) through Project-ID 274200144 – SFB 1227 (projects A06 and A07) and Project-ID 390837967 - EXC 2123, and by the Quantum Valley Lower Saxony Q1 project (QVLS- Q1) through the Volkswagen foundation and the ministry for science and culture of Lower Saxony.

\setcounter{section}{0}  
\section*{Methods}
\renewcommand{\thesection}{M\arabic{section}}
\section{Phase estimation}
    In local phase estimation, assuming that the phase $\phi$ is well centered around a fixed working point $\phi_0$, i.e. $(\phi -\phi_0)^2\ll 1$, and that the estimator $\phi_\mathrm{est}(x,T)$ is locally unbiased, the phase estimation uncertainty is characterized by
    \begin{align}
        (\Delta\phi(T))^2= \sum_x P(x|\phi_0,T) \left[\phi_\mathrm{est}(x,T)-\phi_0\right]^2\label{eq:MSE}
    \end{align}
    with conditional probabilities $P(x|\phi,T) = \tr\left(\ket{x}\bra{x}\Lambda_{\phi,T}[\rho_\mathrm{in}]\right)$. Here, $\{\ket{x}\bra{x}\}$ is the POVM associated with the observable $X$ and corresponding projectors $\ket{x}\bra{x}$ onto the eigenstates with eigenvalues $x$, while $\Lambda_{\phi,T}[\rho_\mathrm{in}]$ denotes the input probe state after free evolution time $T$. The widely established linear estimator reads
    \begin{align}
        \phi_\mathrm{est}(x,T) = \frac{x}{\partial_\phi\braket{X(\phi,T)}|_{\phi=\phi_0}} + \phi_0\label{eq:linearEst}
    \end{align}
    where the scaling factor is given by the inverse of the slope of the signal at the working point. For this particular estimator, the phase estimation uncertainty reads
    \begin{align}
        (\Delta\phi(T))^2 = \frac{(\Delta X(\phi,T))^2}{(\partial_\phi\braket{X(\phi,T)})^2} \Big|_{\phi=\phi_0}\label{eq:deltaphi}
    \end{align}
    with variance $(\Delta X(\phi,T))^2 = \braket{X^2(\phi,T)} -\braket{X(\phi,T)}^2$ of the observable $X$. Without loss of generality, we assumed a vanishing signal $\braket{X(\phi_0,T)}=0$ at the working point (which always can be achieved by appropriately shifting the signal). This expression often is referred to as the method of moments and equivalently can be derived from error propagation. 
    A fundamental limit in the precision of phase estimation for a state $\Lambda_{\phi,T}[\rho_\mathrm{in}]$ is given by the quantum Cram\'{e}r-Rao bound (QCRB)
    \begin{align}
        (\Delta\phi(T))^2 \geq (\Delta\phi(T))_\mathrm{QCRB}^2 = \frac{1}{\mathcal{F}_Q[\Lambda_{\phi,T}[\rho_\mathrm{in}]]}\label{eq:qcrb_general}
    \end{align}
    where $\mathcal{F}_Q[\rho]$ denotes the quantum Fisher information (QFI) of the state $\rho$ and implies the optimization over the measurement and estimation scheme.
    
\section{Noise model}
    Considering spontaneous decay with rate $\Gamma$ and individual dephasing with rate $\gamma$, the dynamics of the system $\Lambda_{\phi,T}$ during the free evolution time $T$ is described by a master equation of the form
    \begin{align}
        \dot{\rho}=-i\omega \left[S_z, \rho\right]+ \frac{\Gamma}{2}\sum_{k=1}^N\mathcal{L}_{S_-^{(k)}}[\rho]  + \frac{\gamma}{2}\sum_{k=1}^N\mathcal{L}_{S_z^{(k)}}[\rho] \label{eq:dynamics}
    \end{align}
    where $S_z^{(k)}=\frac{1}{2}\sigma_z^{(k)}$ and $S_-^{(k)} = \sigma_-^{(k)}$ with single particle Pauli operators $\sigma_j$. The Lindblad superoperators are defined by $\mathcal{L}_A[\rho] := 2 A\rho A^\dag - A^\dag A \rho - \rho A^\dag A$. According to Eq.~\eqref{eq:dynamics}, the phase $\phi$ is imprinted by a rotation around the $z$-axis.

\section{SQL and GHZ Protocols}
    At this point, we provide the relevant expressions, while the corresponding derivations can be found in the Supplementary Materials.
    
    The frequency estimation uncertainty for coherent spin states (CSS) is given by
    \begin{align}
        (\Delta\omega_\mathrm{CSS}(T))^2 = \frac{e^{(\Gamma+\gamma)T}}{N\tau T}
    \end{align}
    and results in the SQL~\eqref{eq:sql} at the optimal interrogation time $T_\mathrm{SQL}^\mathrm{min} = 1/(\Gamma + \gamma)$. Although the parity-GHZ protocol achieves a $N$-times increased sensitivity in the ideal scenario, in the presence of decoherence it collapses faster by the same factor
     \begin{align}
        (\Delta\omega_\mathrm{parity-GHZ}(T))^2 = \frac{e^{(\Gamma + \gamma)NT}}{N^2\tau T}\label{eq:parityGHZ}
    \end{align}
    and thus likewise only performs at the SQL~\eqref{eq:sql}, however, at a $N$-times shorter optimal interrogation time $T_\mathrm{min} = 1/N(\Gamma + \gamma)$. If we perform a correlated measurement according to Eq.~\eqref{eq:GHZGHZ}, the conditional probabilities read
    \begin{multline}
        P\left(x |\phi,T\right)=
        \frac{1}{4}\left[1 + \left(1-e^{-\Gamma T}\right)^N + e^{-\Gamma N T}\right. \\ \mp\left. 2e^{-\frac{\Gamma + \gamma}{2}NT}\cos(N\phi)\right]\label{eq:prob_noerror}
    \end{multline}
        for $x=\pm\frac{N}{2}$ and
    \begin{multline}
        P\left(x |\phi,T\right)=
        \frac{1}{4}\binom{N}{N_-} \left[e^{-\Gamma T(N-N_-)}\left(1-e^{-\Gamma T}\right)^{N_-}\right.\\ +\left. e^{-\Gamma TN_-}\left(1-e^{-\Gamma T}\right)^{N-N_-}\right]\label{eq:prob_error}
    \end{multline}
    for $x\neq\pm\frac{N}{2}$ where $N_- \in \{1,\ldots,N-1\} $ denotes the number of particles in the ground state $\kett{\downarrow}$. Since the phase $\phi$ is solely encoded in the measurement outcomes $x=\pm\frac{N}{2}$, an error detection and mitigation scheme can be implemented as described in the main text by completing the sequence with the highly nonlinear estimator defined in Eq.~\eqref{eq:flag}, resulting in a sensitivity saturating the QCRB~\eqref{eq:GHZQCRB}. If we instead apply a linear estimation rule as introduced in Eq.~\eqref{eq:linearEst}, referred to as `linear-GHZ' protocol, the phase estimation uncertainty is given by
    \begin{multline}
        (\Delta\phi(T)_\mathrm{linear-GHZ})^2 =\\ \frac{e^{(\Gamma+\gamma)NT}}{N^3}\left[1+ (N-1)\left(1 - 2e^{-\Gamma T} + 2e^{-2\Gamma T}\right)\right].\label{eq:linearGHZ}
    \end{multline}
    and thus the QCRB is not saturated for $N>2$. Nevertheless, an improvement compared to the SQL is obtained which, however, vanishes in the asymptotic limit (cf.~\autoref{fig:Theory}B).

\newpage
\ \\
\newpage
\onecolumngrid
\section*{Supplementary Materials}
\setcounter{section}{0}  
\renewcommand{\thesection}{S\arabic{section}}
\setcounter{figure}{0}  
\renewcommand{\thefigure}{S\arabic{figure}}
\setcounter{equation}{0}  
\renewcommand{\theequation}{S\arabic{equation}}

In~\autoref{method:SpinMeasurements} the SQL~\eqref{eq:sql} in the presence of dephasing and spontaneous decay is derived as well as the ultimate lower bound given by Eq.~\eqref{eq:asymptotic}. Properties of the GHZ protocols are obtained in~\autoref{method:GHZprotocols}, namely the evaluation of the QCRB~\eqref{eq:GHZQCRB}, the sensitivities of the parity-GHZ~\eqref{eq:parityGHZ}, linear-GHZ~\eqref{eq:linearGHZ}, heralded-GHZ as well as heralded-uGHZ scheme~\eqref{eq:unbalancedGHZ}, including Eqs.~\eqref{eq:GHZGHZ}, \eqref{eq:flag}, \eqref{eq:unbalancedGHZstate}, \eqref{eq:prob_noerror} and \eqref{eq:prob_error}. The notion of quantum jumps is briefly presented in~\autoref{method:quantumJumps}. In~\autoref{method:asymptotics} the OQI and its asymptotic scaling is discussed.

\section{(Linear) projective spin measurements}\label{method:SpinMeasurements}
Since a rotation of the state around the $z$-axis commutes with the Lindblad operators for spontaneous decay and dephasing, the phase imprint can be separated from the decoherence processes by $\Lambda_{\phi,T}[\rho_\mathrm{in}] = \Lambda_{T,\Gamma,\gamma}[\rho_\phi] $ with $\rho_\phi = \mathcal{R}_\mathbf{z}(\phi) \rho_\mathrm{in}\mathcal{R}_\mathbf{z}^\dagger(\phi)$. Here we introduced the notation $\mathcal{R}_\mathbf{n}(\phi) = \exp(-i\phi S_\mathbf{n})$ of a rotation by the angle $\phi$ around axis $\mathbf{n}$. $S_\mathbf{n}=\mathbf{n}^T\mathcal{S}$ denotes the spin in $\mathbf{n}$-direction with the spin vector $\mathbf{S}=(S_x,S_y,S_z)^T$. Consequently, we have the time evolution according to the decoherence processes governed by
    \begin{align}
        \dot{\rho}_\phi = \mathcal{L}_{T,\Gamma,\gamma}[\rho_\phi] = \frac{\Gamma}{2}\sum_{k=1}^N\mathcal{L}_{S_-^{(k)}}[\rho_\phi]  + \frac{\gamma}{2}\sum_{k=1}^N\mathcal{L}_{S_z^{(k)}}[\rho_\phi].
    \end{align}
        Instead of considering the time evolution of the state $\rho_\phi$, we can equivalently map the decoherence processes to the measurement observable $X$ by exploiting the cyclicity of the trace
        \begin{align}
            \partial_t \braket{X(\phi, T)} = \tr\left(X \dot{\rho}_\phi\right) 
            = \tr\left(X \mathcal{L}_{T,\Gamma,\gamma}[\rho_\phi]\right) 
            = \tr\left(\mathcal{L}_{T,\Gamma,\gamma}^\dagger[X] \rho_\phi\right) 
        \end{align}
        with the adjoint Lindblad operators
        \begin{align}
            \mathcal{L}_{T,\Gamma,\gamma}^\dagger[X] &=  \frac{\Gamma}{2}\sum_{k=1}^N\mathcal{L}_{S_-^{(k)}}^\dagger[X]  + \frac{\gamma}{2}\sum_{k=1}^N\mathcal{L}_{S_z^{(k)}}^\dagger[X] \\
            \mathcal{L}_A^\dagger[X] &:= 2 A^\dag X A - A^\dag A X - X A^\dag A.
        \end{align}
        To determine the phase estimation uncertainty~\eqref{eq:deltaphi}, we have to solve the differential equations for $X$ and $X^2$.
        
        If we consider a measurement of $S_y$, we obtain by direct evaluation and employing properties of the Pauli operators
        \begin{align}
            \braket{S_y(\phi,T)} &= e^{-\frac{\Gamma+\gamma}{2}T}\braket{S_y(\phi,T=0)}\\
            \braket{S_y^2(\phi,T)} &= \frac{N}{4}\left[1-e^{-(\Gamma+\gamma)T}\right] +e^{-(\Gamma+\gamma)T}\braket{S_y^2(\phi,T=0)}.
        \end{align}
        Assuming that the optimal working point is $\phi_0=0$ (otherwise the signal can be shifted accordingly), we obtain
        \begin{align}
            (\Delta\phi(T))^2 =  \frac{\frac{N}{4}\left[e^{(\Gamma+\gamma)T}-1\right] +\braket{S_y^2}}{\braket{S_x}^2}\label{eq:SyMeas}
        \end{align}
        where the expectation values $\langle S_x\rangle$ and $\langle S_y^2 \rangle$ are evaluated with respect to $\rho_\mathrm{in}$.
        
        \paragraph*{CSS ---} The coherent spin state (CSS) $\left(\kett{\uparrow} + \kett{\downarrow}\right)^{\otimes N}/\sqrt{2}^N$ has expectation values $\braket{S_x} = \frac{N}{2}$ and $\braket{S_y^2} = \frac{N}{4}$~\cite{Pezz2018} and thus, the phase estimation uncertainty reads
        \begin{align}
            (\Delta\phi_\mathrm{CSS}(T))^2 =\frac{e^{(\Gamma + \gamma)T}}{N}
        \end{align}
        resulting in the minimal frequency estimation uncertainty given by the SQL~\eqref{eq:sql}, namely $(\Delta\omega_\mathrm{SQL})^2 = \frac{\mathrm{e}\,(\Gamma + \gamma)}{N\tau}$. 
        
        \paragraph*{SSS ---} Squeezed spin states (SSS) generated by OAT have properties~\cite{Kitagawa1993}
        \begin{align}
            \braket{S_x} &= \frac{N}{2}\cos^{N-1}\left(\frac{\mu}{2}\right)\\
            \braket{S_y^2} &= \frac{N}{2}\left\{1+ \frac{1}{4}\left(N-1\right)\left[A - \sqrt{A^2 + B^2}\right]\right\}\\
            A &= 1 - \cos^{N-2}(\mu)\\
            B &= 4\sin\left(\frac{\mu}{2}\right)\cos^{N-2}\left(\frac{\mu}{2}\right)
        \end{align}
        where the twisting strength $\mu$ has to be optimized for each interrogation time $T$ in order to minimize the phase estimation uncertainty~\eqref{eq:SyMeas}.
        
        \paragraph*{Lower bound ---} A lower bound on the frequency estimation uncertainty can be determined following Ref.~\cite{Huelga1997}. With Eq.~\eqref{eq:SyMeas}, the frequency estimation uncertainty~\eqref{eq:freqUncertainty} reads
        \begin{align}
            (\Delta\omega(T))^2 =  \frac{\frac{N}{4}\left[e^{(\Gamma+\gamma)T}-1\right] +\braket{S_y^2}}{T\tau \braket{S_x}^2}.
        \end{align}
        Minimization with respect to the interrogation time $T$ leads to the equation
        \begin{align}
            \frac{N}{4}(\Gamma + \gamma)e^{(\Gamma+\gamma)T_\mathrm{min}}T_\mathrm{min} = \frac{N}{4}\left[e^{(\Gamma+\gamma)T_\mathrm{min}}-1\right] +\braket{S_y^2}
        \end{align}
        and thus we obtain
        \begin{align}
            (\Delta\omega(T_\mathrm{min}))^2 =\frac{(\Gamma + \gamma)e^{(\Gamma + \gamma)T_\mathrm{min}}}{\tau N\left(\frac{\braket{S_x}}{N/2}\right)^2}.
        \end{align}
        Finally, we find the lower bound \eqref{eq:asymptotic}, namely $(\Delta\omega_\mathrm{min}^\mathrm{asym})^2 \geq \frac{\Gamma + \gamma}{N \tau}$, using $e^{(\Gamma + \gamma)T_\mathrm{min}} \geq 1$ and $\braket{S_x} \leq N/2$.

\section{GHZ protocols}\label{method:GHZprotocols}
    After the free evolution time with dynamics according to \eqref{eq:dynamics}, the time evolved GHZ state is given by
    \begin{align}
        \rho_\mathrm{GHZ}(\phi, T) =& \frac{1}{2}\Big(\kett{\downarrow}\braa{\downarrow}^{\otimes N} + e^{-\frac{\Gamma + \gamma }{2}NT}\left[e^{i\phi N}\kett{\downarrow}\braa{\uparrow}^{\otimes N} + e^{-i\phi N}\kett{\uparrow}\braa{\downarrow}^{\otimes N} \right]
        + \left[e^{-\Gamma T}\kett{\uparrow}\braa{\uparrow} + \left(1-e^{-\Gamma T}\right)\kett{\downarrow}\braa{\downarrow}\right]^{\otimes N}\Big).\label{eq:ghz_evolved}
    \end{align}
    \paragraph*{Parity-GHZ protocol ---}
        With the signal of the parity measurement
        \begin{align}
            \braket{\Pi(\phi,T) } = \tr\left(\Pi \rho_\mathrm{GHZ}(\phi, T)\right) = (-1)^N e^{-\frac{\Gamma + \gamma}{2}NT}\cos(N\phi)
        \end{align}
        and $\Pi^2 = \mathds{1}$, the phase estimation uncertainty for the parity measurement $\Pi$ is given by
        \begin{align}
            (\Delta\phi_\mathrm{parity-GHZ}(T))^2 = \frac{e^{(\Gamma + \gamma)NT}}{N^2}
        \end{align}
        at the optimal working point $\phi_0 = \pi/2N$ (since we have a symmetric signal). Although the sensitivity of the parity-GHZ protocol might look superior to the SQL due to the scaling with $1/N^2$, it collapses $N$ times faster and thus, the minimal frequency estimation uncertainty equals the SQL~\eqref{eq:sql} at optimal interrogation time $T_\mathrm{min} =1/N(\Gamma+\gamma)$. Note that $T_\mathrm{min}$ is $N$ times shorter than for the SQL using CSS.

    \paragraph*{QCRB ---}
        To determine the QCRB~\eqref{eq:qcrb_general}, we have to compute the QFI
        \begin{align}
            \mathcal{F}_Q[\rho, S_z] = 2 \sum_{\underset{\lambda_k+\lambda_l > 0}{k,l}}\frac{(\lambda_k-\lambda_l)^2}{\lambda_k + \lambda_l} |\braket{k|S_z|l}|^2\label{eq:qfi_general}
        \end{align}
        where $\lambda_k$ and $\ket{k}$ are the eigenvalues and eigenstates of $\rho$. Here we already used that the phase $\phi$ is imprinted by a rotation around the $z$-axis.
        Fortunately, the time evolved GHZ state~\eqref{eq:ghz_evolved} is already almost diagonal, except for the subspace spanned by the maximal Dicke states $\kett{\uparrow}^{\otimes N}$ and $\kett{\downarrow}^{\otimes N}$. Hence, we only have to diagonalize an effective $2\times 2$-matrix which we denote by
        \begin{align}
            \rho_\mathrm{eff} &= \frac{1}{2}\Big(\left[1 + \left(1 - e^{-\Gamma T}\right)^N\right]\kett{\downarrow}\braa{\downarrow}^{\otimes N} + e^{-\frac{\Gamma + \gamma }{2}NT}\left[\kett{\downarrow}\braa{\uparrow}^{\otimes N} +\kett{\uparrow}\braa{\downarrow}^{\otimes N} \right]+ e^{-\Gamma NT}\kett{\uparrow}\braa{\uparrow}^{\otimes N} \Big)\\
            &= \frac{1}{2}\begin{pmatrix}
                e^{-\Gamma NT} & e^{-\frac{\Gamma + \gamma }{2}NT}\\
                e^{-\frac{\Gamma + \gamma}{2}NT}& 1 + \left(1 - e^{-\Gamma T}\right)^N\\
            \end{pmatrix}.
        \end{align}
        After evaluating the eigensystem of this effective matrix, only one term in Eq.~\eqref{eq:qfi_general} is non-zero and yields the QFI
        \begin{align}
            \mathcal{F}_Q[\rho_\mathrm{GHZ}(\phi, T), S_z] = \frac{2N^2 e^{-(\Gamma + \gamma )NT}}{1 + \left(1-e^{-\Gamma T}\right)^N + e^{-\Gamma N T}}.
        \end{align}
        Finally, we obtain the QCRB~\eqref{eq:GHZQCRB} which leads to an improvement compared to the SQL (cf.~\autoref{fig:Theory}).

    \paragraph*{Heralded-GHZ protocol ---}
        In order to assess the pertinent characteristics of the heralded-GHZ protocol, it is advantageous to employ the identity
        \begin{align}
            \mathcal{U}_\mathrm{GHZ} = \frac{1}{\sqrt{2}}e^{-i\frac{\pi}{4E}}\left[\mathds{1} + i^{N+E}\sigma_x^{\otimes N}\right]
        \end{align}
        with $E=1$ ($E=2$) for $N$ even (odd), which was previously utilized in Ref.~\cite{Leibfried2004}. The derivation of this relation is based on basic algebra in the eigenbasis of $S_x$. Furthermore, it is beneficial to distinguish between two cases, namely $N$ even and odd, as the identity already suggests. Application of $\mathcal{U}_\mathrm{GHZ}$ to the ground state $\kett{\downarrow}^{\otimes N}$ yields
        \begin{align}
            \ket{\psi_\mathrm{in}} = \frac{1}{\sqrt{2}}e^{-i\frac{\pi}{4E}}\left(\kett{\downarrow}^{\otimes N} + i^{N+E}\kett{\uparrow}^{\otimes N}\right).\label{eq:psiGHZ}
        \end{align}
        The exact GHZ state $\ket{\mathrm{GHZ}}=\left(\kett{\uparrow}^{\otimes N} + \kett{\downarrow}^{\otimes N}\right)/\sqrt{2}$ can be obtained by an additional trivial rotation around the $z$-axis, i.e. $\ket{\mathrm{GHZ}}=e^{i\alpha}\mathcal{R}_\mathbf{z}(\theta_E)\mathcal{U}_\mathrm{GHZ}\kett{\downarrow}^{\otimes N}$ where $\theta_E = \frac{\pi}{2N}(N+E)$ and $\alpha = \frac{\pi}{4E}-\frac{\theta_E N}{2}$. The dynamics according to Eq.~\eqref{eq:dynamics} leads to the time evolved state
        \begin{align}
            \rho_\mathrm{in}(\phi,T) &= \frac{1}{2}\Big(\kett{\downarrow}\braa{\downarrow}^{\otimes N} + e^{-\frac{\Gamma + \gamma }{2}NT}\left[e^{i\phi N}(-i)^{N+E}\kett{\downarrow}\braa{\uparrow}^{\otimes N} + e^{-i\phi N}i^{N+E}\kett{\uparrow}\braa{\downarrow}^{\otimes N} \right]\nonumber\\
            &\quad+ \left[e^{-\Gamma T}\kett{\uparrow}\braa{\uparrow} + \left(1-e^{-\Gamma T}\right)\kett{\downarrow}\braa{\downarrow}\right]^{\otimes N}\Big).
        \end{align}
        Applying the measurement transformation $\mathcal{U}_\mathrm{GHZ}$ of the observable $X$ \eqref{eq:GHZGHZ} to the time evolved state results in the final state
        \begin{align}
            \rho_\mathrm{final}(\phi,T) &= \mathcal{U}_\mathrm{GHZ}^\dagger \, \rho_\mathrm{in}(\phi,t)\,  \mathcal{U}_\mathrm{GHZ}\\
            &= \frac{1}{4}\Big\{\kett{\downarrow}\braa{\downarrow}^{\otimes N} +i^{N+E}\kett{\downarrow}\braa{\uparrow}^{\otimes N} + (-i)^{N+E}\kett{\uparrow}\braa{\downarrow}^{\otimes N} + \kett{\uparrow}\braa{\uparrow}^{\otimes N}\\
            &\quad+ e^{-\frac{\Gamma + \gamma}{2}NT}\Big[e^{iN\phi }(-i)^{N+E}\Big(\kett{\downarrow}\braa{\uparrow}^{\otimes N} + (-i)^{N+E}\kett{\uparrow}\braa{\uparrow}^{\otimes N} + i^{N+E}\kett{\downarrow}\braa{\downarrow}^{\otimes N}+\kett{\uparrow}\braa{\downarrow}^{\otimes N} \Big)\\
            &\quad+e^{-iN\phi}i^{N+E}\Big(\kett{\uparrow}\braa{\downarrow}^{\otimes N} +(-i)^{N+E}\kett{\downarrow}\braa{\downarrow}^{\otimes N} + i^{N+E}\kett{\uparrow}\braa{\uparrow}^{\otimes N} + \kett{\downarrow}\braa{\uparrow}^{\otimes N}\Big) \Big]\\
            &\quad+\left[e^{-\Gamma T}\kett{\uparrow}\braa{\uparrow} + \left(1-e^{-\Gamma T}\right)\kett{\downarrow}\braa{\downarrow}\right]^{\otimes N}
            + i^{N+E}\left[e^{-\Gamma T}\kett{\downarrow}\braa{\uparrow} + \left(1-e^{-\Gamma T}\right)\kett{\uparrow}\braa{\downarrow}\right]^{\otimes N}\\
            &\quad+ (-i)^{N+E}\left[e^{-\Gamma T}\kett{\uparrow}\braa{\downarrow} + \left(1-e^{-\Gamma T}\right)\kett{\downarrow}\braa{\uparrow}\right]^{\otimes N} 
            + \left[e^{-\Gamma T}\kett{\downarrow}\braa{\downarrow} + \left(1-e^{-\Gamma T}\right)\kett{\uparrow}\braa{\uparrow}\right]^{\otimes N}\Big\} .\label{eq:ghzghz_final}
        \end{align}
        On $\rho_\mathrm{final}$ a projective measurement of $S_z$ is performed. We will refer to this interrogation scheme combined with the standard linear estimator as `linear-GHZ' protocol. The signal reads
        \begin{align}
            \braket{X(\phi,T)} = \tr\left(X \rho_\mathrm{in}(\phi, T)\right) =  \tr\left(S_z \rho_\mathrm{final}(\phi, T)\right) 
            =-\frac{N}{2}e^{-\frac{\Gamma + \gamma }{2}NT}\cos(N\phi)
        \end{align}
        and the second moment is given by
        \begin{align}
            \braket{X^2(\phi,T)} = \frac{N}{4}\left[1+ (N-1)\left(1 - 2e^{-\Gamma T} + 2e^{-2\Gamma T}\right)\right]
        \end{align}
        where we used $S_z^2 = \frac{N}{4}\mathds{1} + \frac{1}{4}\sum_{j\neq k} \sigma_z^{(j)} \sigma_z^{(k)}$. Again, we have a symmetric signal with optimal working point $\phi_0 = \frac{\pi}{2N}$. Consequently, the phase estimation uncertainty reads
        \begin{align}
            (\Delta\phi_\mathrm{linear-GHZ}(T))^2 = \frac{e^{(\Gamma+\gamma)NT}}{N^3}\left[1+ (N-1)\left(1 - 2e^{-\Gamma T} + 2e^{-2\Gamma T}\right)\right]
        \end{align}
        and thus yields a lower phase estimation uncertainty than the SQL. For $N=2$ the linear-GHZ protocol saturates the QCRB. However, for $N>2$ the QCRB is not saturated and the SQL is asymptotically approximated (cf.~\autoref{fig:Theory}B).
        
        The conditional probabilities can be directly inferred from the final state~\eqref{eq:ghzghz_final} and read
        \begin{align}
            P\left(x |\phi,T\right) = \begin{cases}
                \frac{1}{4}\left[1 + \left(1-e^{-\Gamma T}\right)^N + e^{-\Gamma N T} \mp 2e^{-\frac{\Gamma + \gamma}{2}NT}\cos(N\phi)\right] &\text{if }x=\pm\frac{N}{2} \\
                \frac{1}{4}\binom{N}{N_-} \left[e^{-\Gamma T(N-N_-)}\left(1-e^{-\Gamma T}\right)^{N_-} + e^{-\Gamma TN_-}\left(1-e^{-\Gamma T}\right)^{N-N_-}\right] &\text{if }x=\frac{N}{2}-N_- \\
            \end{cases}\label{eq:condProbHeralded}
        \end{align}
        where $N_- \in \{1,\ldots,N-1\} $ denotes the number of particles in the ground state $\kett{\downarrow}$. It is important to note that the phase information is solely encoded in the measurement of the maximal Dicke states, i.e. $x=\pm\frac{N}{2}$, which has profound implications. Only if we measure $x=\pm\frac{N}{2}$ we get some information on the phase while for $x\neq \pm\frac{N}{2}$ we basically measure noise with a binomial(-like) distribution characterized by the spontaneous decay rate $\Gamma$ and interrogation time $T$. Consequently, using the linear estimator~\eqref{eq:linearEst} we estimate randomly based on this noise distribution. Metaphorically speaking this corresponds to a blind guess and spoils the sensitivity. Furthermore, the measurement outcomes can be interpreted as a flag for spontaneous decay. If none of the particles decayed, the time evolved state accumulated a relative phase and effectively one of the two GHZ-like states $\mathcal{U}_\mathrm{GHZ}\kett{\downarrow}^{\otimes N}$ or $\mathcal{U}_\mathrm{GHZ}\kett{\uparrow}^{\otimes N}$ is measured. However, if a particle decayed, the subspace spanned by the maximal Dicke states is left and we obtain a measurement outcome $x\neq \pm\frac{N}{2}$ depending on the number of particles that decayed. We will take advantage of both observations by introducing a dedicated estimator~\eqref{eq:flag} which discards all measurement results but the maximal ones, i.e. $x=\pm\frac{N}{2}$. In this case, phase estimation uncertainty according to Eq.~\eqref{eq:MSE} reads
        \begin{align}
            (\Delta\phi(T))^2 = \frac{N^2}{4}\frac{P\left(x=+\frac{N}{2} |\phi_0,T\right)+P\left(x=-\frac{N}{2} |\phi_0,T\right)}{\left(\partial_\phi \braket{X(\phi,T)}|_{\phi=\phi_0}\right)^2}
        \end{align}
        und finally results in the QCRB~\eqref{eq:GHZQCRB}, i.e. $(\Delta\phi_\mathrm{QCRB}^\mathrm{GHZ}(T))^2  = \frac{ \mathrm{e}^{(\Gamma + \gamma )NT}}{2N^2} \left[1 + e^{-\Gamma N T}+\left(1-e^{-\Gamma T}\right)^N\right]$. Note that the application of the nonlinear estimator alternatively can be imitated by the designed observable $\tilde{X} = \frac{N}{2}\mathcal{U}_\mathrm{GHZ} \left(\ket{\uparrow}\bra{\uparrow}^{\otimes N} -  \ket{\downarrow}\bra{\downarrow}^{\otimes N}\right)\mathcal{U}_\mathrm{GHZ}^\dagger$. Interestingly, the POVM associated with $\tilde{X}$ and the symmetric logarithmic derivative (SLD) $ L_{\Lambda_{\phi, T}[\rho_\mathrm{GHZ}]}$ coincide. In our framework, $L_\rho$ is implicitly defined via $\frac{1}{2}\{L,\rho\}=-i[S_z,\rho]$ and gives rise to the QFI $\mathcal{F}_Q[\rho]=\tr\left(\rho L^2\right)$.

    \paragraph*{Heralded-uGHZ protocol ---} The initial state \eqref{eq:unbalancedGHZstate} is obtained by applying the rotation $\mathcal{R}_z(\theta)$ and OAT interaction $\mathcal{U}_\mathrm{GHZ}$ to the state \eqref{eq:psiGHZ}
    \begin{align}
        \ket{\psi_\mathrm{in}} = \mathcal{U}_\mathrm{GHZ} \mathcal{R}_z(\theta) \mathcal{U}_\mathrm{GHZ} \kett{\downarrow}^{\otimes N}= \alpha_\theta\kett{\downarrow}^{\otimes N}+\beta_\theta\kett{\uparrow}^{\otimes N}  = e^{-i\frac{\pi}{2E}} \left[i\sin\left( \tfrac{\theta N}{2}\right)\kett{\downarrow}^{\otimes N}+ i^{N+E}\cos\left( \tfrac{\theta N}{2}\right)\kett{\uparrow}^{\otimes N}\right].\label{eq:initial_state}
    \end{align}
    Fortunately, the remaining evaluation is analogous to the heralded-GHZ protocol but with modified coefficients $\alpha_\theta$ and $\beta_\theta$ of the maximal Dicke states $\kett{\downarrow}^{\otimes N}$ and $\kett{\uparrow}^{\otimes N}$. The optimal rotation angle reads
    \begin{align}
        \theta = \frac{2}{N} \arctan\left(\sqrt[4]{e^{-N\Gamma T} + \left(1-e^{-\Gamma T}\right)^N}\right) \label{eq:optimal_angle}
    \end{align}
    and the associated minimal phase estimation uncertainty is given by \eqref{eq:unbalancedGHZ}, i.e. $(\Delta\phi(T))^2 = \frac{e^{(\Gamma + \gamma)N T}}{4N^2 } \left[1 + \sqrt{e^{-N\Gamma T} + \left(1-e^{-\Gamma T}\right)^N}\right]^2$. Moreover, the corresponding initial state \eqref{eq:initial_state} has populations
    \begin{align}
        P_\mathrm{in}\left(-\frac{N}{2}\right) &= \tr\left(\kett{\downarrow}\braa{\downarrow}^{\otimes N} \ket{\psi_\mathrm{in}}\bra{\psi_\mathrm{in}}\right) = \frac{\sqrt{e^{-N\Gamma T} + \left(1-e^{-\Gamma T}\right)^N}}{1+\sqrt{e^{-N\Gamma T} + \left(1-e^{-\Gamma T}\right)^N}} \\
        P_\mathrm{in}\left(+\frac{N}{2}\right) &= \tr\left(\kett{\uparrow}\braa{\uparrow}^{\otimes N} \ket{\psi_\mathrm{in}}\bra{\psi_\mathrm{in}}\right) = \frac{1}{1+\sqrt{e^{-N\Gamma T} + \left(1-e^{-\Gamma T}\right)^N}}
    \end{align}
    and consequently indeed represents a non-equal superposition, with a higher population in the excited state. Nevertheless, again the phase is only encoded in the maximal Dicke states after the free evolution time and the final OAT interaction
    \begin{align}
        P\left(x |\phi,T\right) = \begin{cases}
            \frac{1}{2}\sqrt{e^{-N\Gamma T} + \left(1-e^{-\Gamma T}\right)^N}  \pm e^{-\frac{\Gamma +\gamma}{2}NT}\frac{\sqrt[4]{e^{-N\Gamma T} + \left(1-e^{-\Gamma T}\right)^N}}{1+\sqrt{e^{-N\Gamma T} + \left(1-e^{-\Gamma T}\right)^N}}\sin(N\phi) &\text{if }x=\pm\frac{N}{2} \\
            \frac{1}{2}\frac{1}{1+\sqrt{e^{-N\Gamma T} + \left(1-e^{-\Gamma T}\right)^N}}\binom{N}{N_-}\left[e^{-\Gamma T (N- N_-)}\left(1-e^{-\Gamma T}\right)^{N_-} + e^{-\Gamma T N_-}\left(1-e^{-\Gamma T}\right)^{N-N_-}\right] &\text{if }x=\frac{N}{2}-N_- \\
        \end{cases}
    \end{align}
    and thus suggests the use of the dedicated estimator \eqref{eq:flag}.\\
    \ \\
    Note that the optimal measurement and estimation schemes for the heralded-GHZ and heralded-uGHZ protocol were discovered by a systematic study of a variational class, similar to Refs.~\cite{Kaubruegger2021,Thurtell2022}.

\section{Quantum jumps}\label{method:quantumJumps}

    In the following we will discuss the notion of quantum jumps in open quantum systems interacting with their environment based on master equations. In particular, we aim to identify and explain the `no-jump dynamics' which is crucial for the presented GHZ protocols. The dynamics based on a general master equation is governed by
    \begin{align}
        \dot{\rho} = -i[H,\rho] + \left(A\rho A^\dagger - \frac{1}{2}A^\dagger A \rho - \frac{1}{2}\rho A^\dagger A \right)
    \end{align}
    with Hamiltonian $H$ and Lindblad (jump) operator $A$ representing an arbitrary dissipative process (quantum jump). For simplicity, we will consider a single jump operator, although the method can easily be generalized to include several noise processes. In our framework, the Hamiltonian $H=\omega S_z$ is given by the frequency deviation $\omega$ and spin operator along the $z$-direction $S_z$, while the jump operators for each qubit are $A = \sqrt{\Gamma}\sigma_-$ and $A = \frac{\sqrt{\gamma}}{2}\sigma_z$ for spontaneous decay and individual dephasing, respectively. It is instructive to rewrite the master equation according to
    \begin{align}
        \dot{\rho} = -i(H_\mathrm{eff}\rho -\rho H_\mathrm{eff}^\dagger) + A\rho A^\dagger \qquad \text{with}\qquad H_\mathrm{eff} = H - \frac{i}{2}A^\dagger A.\label{eq:me_eff}
    \end{align}
   The dynamics can be interpreted as a combination of a continuous  and a stochastic evolution. The continuous evolution, also called `no-jump dynamics', is determined by the effective Hamiltonian $H_\mathrm{eff}$ and does not imply any quantum jumps, i.e. no quanta are exchanged with the environment. Nevertheless, an effective decay is caused by the non-hermitian part of $H_\mathrm{eff}$. In contrast, the stochastic contribution is characterized by the term $A\rho A^\dagger$ and results in sudden quantum jumps, interrupting the continuous evolution. These jumps reflect discrete events like the emission of a photon.

    A series expansion of the formal solution to the the master equation \eqref{eq:me_eff} is given by
    \begin{align}
        \rho(t,t_0) = G(t,t_0)\rho(t_0) + \int_{t_0}^{t}\dd t_1 G(t,t_1)JG(t_1,t_0)\rho(t_0) + \int_{t_0}^{t}\dd t_2 \int_{t_0}^{t}\dd t_1 G(t,t_2)JG(t_2,t_1)JG(t_1,t_0)\rho(t_0) + \ldots \label{eq:series}
    \end{align}
    with superoperators $G(t,t_0)B = e^{-iH_\mathrm{eff}(t-t_0)}Be^{iH_\mathrm{eff}^\dagger(t-t_0)}$ and $JB = A B A^\dagger$ representing the continuous non-unitary time propagation and the quantum jump, respectively. The terms in Eq.~\eqref{eq:series} include an increasing number of jumps (characterized by the number of $J$ operators) in the time interval $[t_0,t]$. In particular, the first term $G(t,t_0)\rho(t_0)$ does not include a single jump and thus is referred to as `no-jump dynamics'.

\section{Optimal quantum interferometer}\label{method:asymptotics}
    
    The optimal quantum interferometer (OQI) defines the ultimate lower limit in phase estimation for a particular setup. Considering a phase evolution implemented by an unitary rotation $\mathcal{R}_z(\phi)$, the OQI $(\Delta\phi_\mathrm{OQI}(T))^2 = 1/\mathcal{F}_\mathrm{OQI}$ can be determined by optimizing the QFI over all input states
    \begin{align}
        \mathcal{F}_\mathrm{OQI} = \sup_{\ket{\psi}} \mathcal{F}_Q\left[\Lambda_T[\ket{\psi}\bra{\psi}],S_z\right]
    \end{align}
    where $\Lambda_T$ describes any decoherence processes. In an ideal scenario, i.e.  $\Lambda_T[\ket{\psi}\bra{\psi}] = \ket{\psi}\bra{\psi}$, the well known Heisenberg limit $\mathcal{F}_\mathrm{OQI}=N^2$ is obtained. However, no general expressions for the OQI exist in the presence of decoherence and a numerical evaluation is necessary. 
    \paragraph*{Numerical computation ---} Considering permutational invariance of the ensemble, a basis is given by the $n_\mathrm{PI}=\left(\frac{N}{2}+1\right)^2-\frac{1}{4}\mathrm{mod}_2(N)$ permutationally invariant Dicke states, resulting in a considerably smaller subspace $\mathcal{O}(N^2)$ compared to the full $2^N$-dimensional Hilbert space. Additionally assuming a pure initial state with total spin $S=N/2$, $\ket{\psi}$ contains $N+1$ non-zero coefficients at most. Nevertheless, a brute force optimization is very inefficient. A more efficient approach to compute the OQI is presented in Ref.~\cite{Macieszczak2013}, which iteratively optimizes the initial probe state and the observable. Starting from an arbitrary state, repeatedly the optimal measurement (via the QFI) and the corresponding optimal probe state are determined. Both involve the time evolution with $\Lambda_T$ and a matrix diagonalization. Fortunately, the dynamics for a fixed interrogation time is identical in each step and thus can be evaluated once in advance. Accordingly, the computational complexity of the time evolution in each step can be neglected. Since the permutationally invariant subspace has dimension $\mathcal{O}(N^2)$, matrix diagonalization of the state and observable has complexity $\mathcal{O}(N^6)$. Note, however, that in practice, the complexity might be reduced due to the block-diagonal structure of the states and spin operators, resulting in only $\mathcal{O}(N^3)$ of the $\mathcal{O}(N^4)$ elements being non-zero. Hence, the complexity of a single step in the iterative procedure is $\mathcal{O}(N^{6})$ at most. Consequently, despite its polynomial scaling, the iterative algorithm becomes challenging in terms of computational time with increasing ensemble size $N$. Additionally, the choice of the first state in the algorithm might affect convergence and result in local extrema. Moreover, numerical errors increase with $N$. Altogether, the evaluation of the OQI and minimization with respect to the interrogation time is only feasible for a few tens of qubits on a standard PC in the presence of decoherence.

    \paragraph*{Asymptotic scaling ---}In contrast,  in the asymptotic limit ($N \gg 1$) lower bounds on the sensitivity of the OQI can be derived for particular decoherence processes. In Ref.~\cite{Huelga1997} a lower bound for individual dephasing was derived that yields a maximal improvement of $1/e$ over the SQL, considering arbitrary input states and a projective measurement of the spin in a suitable direction. The extension to spontaneous decay results in the same improvement, and the lower limit is given by Eq.~\eqref{eq:asymptotic}, derived in~\autoref{method:SpinMeasurements}. Interestingly, this already represents the asymptotic ultimate lower bound as derived in Refs.~\cite{Escher2011,Demkowicz_Dobrza_ski_2012,DemkowiczDobrzaski2014,Knysh2014} where the tightness was additionally proved. Generalizing the investigations of Ref.~\cite{UlamOrgikh2001} to additionally include spontaneous decay (cf.~\autoref{method:SpinMeasurements}), Eq.~\eqref{eq:asymptotic} is asymptotically saturated by squeezed spin states (SSS), e.g. generated by OAT interactions. Note that in the presence of spontaneous emission a lower bound smaller by a factor of $4$ compared to Eq.~\eqref{eq:asymptotic} was derived in Refs.~\cite{Demkowicz_Dobrza_ski_2012,DemkowiczDobrzaski2014} which might be achievable by means of additional ancilla systems or adaptive quantum feedback strategies as quantum error correction schemes. However, saturability remains an open question and is beyond the scope of this work. Refs. \cite{Sekatski2017,Kurdziaek2023} give a comprehensive overview over different quantum control strategies and corresponding bounds. 
\end{document}